\documentclass[12pt,a4paper]{article}
\usepackage{amsmath}
\usepackage{amsxtra}
    \usepackage{amstext}
    \usepackage{amssymb}
    \usepackage{latexsym}
    \usepackage{graphicx}
\usepackage{color}
\usepackage{graphics}

\newcommand{\p}{\vspace{6pt}\noindent}

\advance\textheight by \topskip
\makeatletter

\@addtoreset{equation}{section}
\def\section{\@startsection {section}{1}{\z@}{-8.5ex plus -1ex minus
 -.2ex}{3.3ex plus .2ex}{\large\bf}}
\def\subsection{\@startsection{subsection}{2}{\z@}{-3.25ex plus
 -1ex minus -.2ex}{1.5ex plus .2ex}{\bf}}
\def\subsubsection{\@startsection{subsubsection}{3}{\z@}{-3.25ex plus%
 -1ex minus -.2ex}{1.5ex plus .2ex}{\sl}}

\begin{document}

\begin{titlepage}
\vspace*{-2cm}
\begin{flushright}
\end{flushright}

\vspace{0.3cm}

\begin{center}
{\Large {\bf The classical nonlinear Schr\"odinger model with a new integrable boundary}} \vspace{1cm} {\Large {\bf }}\\
\vspace{1cm}
{\large C.\ Zambon\footnote{\noindent E-mail: {\tt cristina.zambon@durham.ac.uk}}} \\
\vspace{0.3cm}
{\em Department of Physics \\ Durham University, Durham DH1 3LE, U.K.} \\

\vspace{2cm} {\bf{ABSTRACT}}\\

\vspace{.5cm}

\end{center}

\p  A new integrable boundary for the classical nonlinear Schr\"odinger model is derived by dressing a boundary with a defect. A complete investigation of the integrability of the new boundary is carried out in the sense that the boundary ${\cal K}$ matrix is derived and the integrability is proved via the classical $r$-matrix. The issue of conserved charges is also discussed. The key point in proving the integrability of the new boundary is the use of suitable modified Poisson brackets. Finally, concerning the kind of defect used in the present context, this investigation offers the opportunity to prove - beyond any doubts - their integrability.

\vfill
\end{titlepage}

\section{Introduction}

\p The classical nonlinear Schr\"odinger (NLS) model with integrable boundaries has been investigated by Sklyanin in \cite{Sklyanin88}. In his seminal paper, he also provided a systematic approach to the study of boundary conditions compatible with integrability.
The purpose of this article is to provide a new integrable boundary associated with a non constant ${\cal K}$ matrix, unlike the one found in \cite{Sklyanin88}. The existence of a non constant boundary ${\cal K}$ matrix is not a new phenomenon. It is worth remembering the examples in \cite{dfr2008} for the NLS model and in \cite{bd2001} for the sine-Gordon model. However, all these examples have in common the feature that the dynamical part of the ${\cal K}$ matrix is represented by additional degrees of freedom present at the boundary location. In other words, the NLS fields and the sine-Gordon field, respectively, do not appear in the ${\cal K}$ matrix.

\p The boundary presented in the present article does not have additional dynamical variables and the ${\cal K}$ matrix is characterized by the presence of the NLS fields. Such a new boundary is obtained by dressing a Dirichlet boundary with a type I defect \cite{cz06NLS}. The idea to dress a boundary with a defect in order to obtain new boundaries is not new. Such a technique has been used before in \cite{bajnok07}, \cite{bu2008} and \cite{cz2012}. However, it is the first time it is applied to the NLS model. It is interesting to point out that the specific dressing taken into account in this article is not possible in the case of the sine-Gordon model, in the sense that it does not lead to a new boundary. In fact, a type II defect must be used instead \cite{cz2012}. On the other hand, it is possible in the case of the complex sine-Gordon model \cite{bu2008}. It is worth reminding that the type I defect - first introduced in \cite{bczlandau} in the context of the sine-Gordon model - manifests itself as a discontinuity in the field at a specific location. Amongst its many features, there is the possibility to retain a generalised momentum and the fact that the defect conditions have the form of B\"acklund transformation fixed at the defect location. Not all integrable models can support a type I defect. Amongst the single scalar filed theory that cannot support such a defect it is worth mentioning the Tzitz\'eica model - also known as the $a^{(2)}_2$ affine Toda model - for which a type II defect has been introduced in \cite{cz2009}. The main difference with respect to the type I defect is that the type II requires the presence of extra degrees of freedom at the defect location. In the case of the NLS model, see also \cite{ad2012NLS} for different types of defects.

\p In this article the Lax pair technique will be used in conjunction with the classical $r$-matrix formalism in order to prove the integrability of the new boundary. In this context, it will be shown how the standard Poisson brackets need to be modified in order to take into account the non trivial coupling of the fields at the boundary. It will be clear that this modification is required by the presence of time derivatives in the boundary potential appearing in the Lagrangian density. The same technique can also be applied to the defect case and this allows to prove, beyond any doubts, the integrability of the kind of defect used in this article. In fact, it is worth reminding that a complete proof of the defect integrability via the classical $r$-matrix was missing so far, despite the fact that strong evidence has been collected to support this statement. On this issue, see \cite{bczlandau}, \cite{cz06NLS}, \cite{cz2009}, \cite{czquantum} and also \cite{Caud08}, \cite{HabKun07}, \cite{ad2012}.

\section{The nonlinear Schr\"odinger model in the bulk}
\label{section2}

\p The nonlinear Schr\"odinger equation with a cubic attractive interaction term will be taken to be defined by the field equation
\begin{equation}\label{NLSequation}
iu_t+u_{xx}+2(u\bar{u})u=0.
\end{equation}
This may be derived in the bulk using an action principle based on the Lagrangian density
\begin{equation}\label{LagrangianBulk}
{\cal
L}=\frac{i}{2}(\bar{u}u_t-\bar{u}_tu)-|u_x|^2+|u|^4.
\end{equation}

\p In order to investigate the integrability of the NLS with boundary both a Lax pair and a $r$-matrix approach will be used \cite{FaddeevTakh86}. It is therefore necessary to introduce the main ingredients of the formulation and to fix the notation. The Lax pair is
\begin{eqnarray}\label{NLSLaxPair}
U&=&i\bar{u}\sigma_++iu\sigma_-+\frac{\lambda}{2i}\sigma_3,\nonumber\\
V&=&-\left(i|u|^2+\frac{\lambda^2}{2i}\right)\sigma_3+(\bar{u}_x-i\lambda \bar{u})\sigma_+-(u_x-i\lambda u)\sigma_-,
\end{eqnarray}
where
$$\sigma_+=\left(
\begin{array}{cc}
            0 & 1 \\
            0 & 0
          \end{array}
          \right),\quad
\sigma_-=\left(\begin{array}{cc}
            0 & 0 \\
            1 & 0
          \end{array} \right),\quad
\sigma_3=\left(\begin{array}{cc}
            1 & 0 \\
            0 & -1
          \end{array} \right),
$$
with the property
\begin{equation}\label{zerocurvaturecondition}
\partial_tU-\partial_xV+[U,V]=0 \Longleftrightarrow [\partial_t-V,\partial_x-U]=0,
\end{equation}
which holds for any choice of the spectral parameter $\lambda$.
Note that the Lax pair is defined up to a gauge transformation ${\cal G}$, that is
\begin{eqnarray}\label{gaugetranformations}
\partial_x {\cal G}=\tilde{U}{\cal G}-{\cal G}U,\nonumber\\
\partial_t {\cal G}=\tilde{V}{\cal G}-{\cal G}V,
\end{eqnarray}
where $(\tilde{U},\tilde{V})$ is an equivalent Lax pair.

\p The zero curvature condition \eqref{zerocurvaturecondition} with matrices \eqref{NLSLaxPair} implies the NLS equation \eqref{NLSequation}. It represents a compatibility condition for the following
overdetermined system of linear equations
\begin{eqnarray}\label{linearsystem}
\partial_x T(x,y,\lambda)&=&U(x,\lambda)T(x,y,\lambda),\nonumber\\
\partial_t T(x,y,\lambda)&=&V(x,t)T(x,y,\lambda)-T(x,y,\lambda)V(y,t),
\end{eqnarray}
where $T$ represents a solution of the system with the initial condition
$$T(x,x,\lambda)=I.$$
This is called the monodromy matrix and it can be represented as the following path-ordered exponential
\begin{equation}
T(x,y,t,\lambda)={\cal P}\left(\exp\int_y^x dz\, U(z,t,\lambda)\right).
\end{equation}
Some useful properties of the monodromy matrix are:
\begin{eqnarray}\label{Tproperties}
\partial_y T(x,y,\lambda)&=&-T(x,y,\lambda)U(y,\lambda),\nonumber\\
T(x,y,\lambda)&=&T(x,w,\lambda)T(w,y,\lambda),\nonumber\\
T(x,y,\lambda)&=&T^{-1}(y,x,\lambda),\nonumber\\
\bar{T}(x,y,\lambda)&=&\sigma_2T(x,y,\bar{\lambda})\sigma_2,\nonumber\\
\mbox{det}\,T(x,y,\lambda)&=&1.
\end{eqnarray}

\p The standard definition of the Poisson brackets, which holds in the case of the NLS in the bulk, is:
\begin{equation}\label{PoissonBracketsbulk}
\{A , B\}=i\int_{-\infty}^{\infty}dz \left(\frac{\delta A}{\delta u(z)} \frac{\delta B}{\delta \bar{u}(z)}-\frac{\delta A}{\delta \bar{u}(z)} \frac{\delta B}{\delta u(z)}\right),
\end{equation}
where $A$ and $B$ are two observables, functionals of the fields $u$ and $\bar{u}$. The Poisson brackets for the NLS fields are:
$$\{u(x),u(y)\}=\{\bar{u}(x),\bar{u}(y)\}=0,\quad \{u(x),\bar{u}(y)\}=i\delta(x-y).$$
It follows that the Poisson brackets between two monodromy matrices for any choices of the spectral parameters $\lambda$ and $\mu$ are:
\begin{equation}\label{PBforT}
\{T_1(x,y,\lambda), T_2(x,y,\mu)\}=[r_{12}(\lambda-\mu),T_1(x,y,\lambda) T_2(x,y,\mu)],
\end{equation}
where
$$T_1(\lambda)=T(\lambda)\otimes I,\quad T_2(\mu)=I \otimes T(\mu).$$
The classical $r$-matrix for the NLS model $r_{12}$ is
\begin{equation}\label{rmatrix}
r_{12}(\lambda)=\frac{P}{\lambda},
\end{equation}
where $P$ is the permutation matrix in $\mathbb{C}^2\otimes \mathbb{C}^2$, namely
$$P=\left(
\begin{array}{cccc}
  1 & 0 & 0 & 0 \\
  0 & 0 & 1 & 0 \\
  0 & 1 & 0 & 0 \\
  0 & 0 & 0 & 1
\end{array}
\right).$$
Note that
\begin{equation}\label{permutationoperator}
P(A\otimes B)= (B\otimes A)P,
\end{equation}
where $A$ and $B$ are $2\times 2$ matrices.

\p Finally, a generating functional for the conserved charges is defined by
$Q(\lambda)=\mbox{tr}\, T(\infty, -\infty;\lambda).$ Provided the fields and their derivatives satisfy suitable conditions at $\pm\infty$, that is they decay to zero\footnote{In particular, they satisfy Schwartz boundary conditions.},
it can be shown that
$\{Q(\lambda),Q(\mu)\}=0.$
As a consequence, the conserved charges are in involution.

\section{The nonlinear Schr\"odinger model with an integrable defect}

\p Some years ago is was shown in \cite{cz06NLS} how a defect can be incorporated into the NLS model without spoiling integrability. One of the key features of this kind of defect is the existence of a generalized momentum that remains conserved despite the presence of a defect at a fixed location. In order to show how integrability is preserved, an argument based on the modified Lax pair was provided. In particular, a generating functional for the conserved charges was constructed, though the fact that the charges are in involution was not proved. In other words two generating functionals $Q(\lambda)$ and $Q(\mu)$ were not proved to Poisson commute. For the time being it is assumed that the integrability argument developed in \cite{cz06NLS} suffices, that is that the defect in \cite{cz06NLS} is indeed integrable. For convenience and to fix notation, its key features will be presented in this section.

\p Suppose that a defect is located at $x=0$. The field to either side of it will be denoted $v$ and $u$\footnote{Note that the fields $u$ and $v$ are interchanged with respect to \cite{cz06NLS}}.
A defect contribution ${\bf D}$ at $x=0$ will be added and the full Lagrangian will be
$$
L=\int_{-\infty}^0 dx\,{\cal
L}_v+{\bf D}+\int_0^{\infty} dx\,{\cal
L}_u.$$
The corresponding defect conditions at $x=0$ are:
\begin{eqnarray}\label{defectconditions0}
v_x&=&\phantom{-}\frac{\partial{\bf D}}{\partial\bar{v}}-\frac{\partial}{\partial
t}\frac{\partial{\bf
D}}{\partial\bar{v}_t}\nonumber\\
u_x&=&-\frac{\partial{\bf D}}{\partial\bar{u}}+\frac{\partial}{\partial
t}\frac{\partial{\bf D}}{\partial\bar{u}_t},
\end{eqnarray}
A suitable defect contribution, which allows a generalized momentum to remain conserved was found to be
\begin{equation}
{\bf D}=\frac{i\Omega}{2}\left(\frac{v_t-u_t}{v-u}-\frac{\bar{v}_t-\bar{u}_t}{\bar{v}-\bar{u}}\right)+\Omega\,{\cal D},
\end{equation}
with
\begin{equation}\label{defectpotential}
\Omega=(\alpha^2-|v-u|^2)^{1/2},\quad {\cal D}=\frac{\Omega^2}{3}+(|v|^2+|u|^2),
\end{equation}
and $\alpha$ a real parameter. The term $ {\cal D}$ will be called the defect potential.
With this choice for the defect contribution, the defect conditions \eqref{defectconditions0} at $x=0$ become
\begin{eqnarray}\label{defectconditions}
v_x=-\frac{1}{2}\left(\frac{i(v_t-u_t)}{\Omega}-(v+u)\Omega+
\frac{(v-u)(|v|^2+|u|^2)}{\Omega}\right),\nonumber\\
u_x=-\frac{1}{2}\left(\frac{i(v_t-u_t)}{\Omega}+(v+u)\Omega+
\frac{(v-u)(|v|^2+|u|^2)}{\Omega}\right).
\end{eqnarray}
In \cite{cz06NLS} it was noticed as these defect conditions represent B\"acklund transformations fixed at $x=0$. Actually, making use of this fact a slightly more general defect contribution was found in \cite{Caud08}. In that case another real parameter $\beta$ is added and the defect potential $ {\cal D}$ is modified as follows
$${\cal D}=\frac{\Omega^2}{3}+(|v|^2+|u|^2) -\Omega \beta ^2+i\beta (\bar{v}u-\bar{u}v).$$
The defect conditions are also modified accordingly.

\p The generating functional for the conserved charges is:
\begin{equation}\label{defectGeneratingfunctional}
{\cal Q}_D(\lambda)=\mbox{tr}\,{\cal T}_D(\lambda)
\end{equation}
where the monodromy matrix ${\cal T}_)D$ is:
$${\cal T}_D(\lambda)=T(\infty, 0,t;\lambda)\,{\cal K}_D(\lambda)\,
T(0,-\infty,t;\lambda),$$
with
\begin{equation}\label{Kmatrix}
{\cal K}_D=\left(1+\frac{\beta}{\lambda}\right)I+\frac{1}{\lambda}\left((v-u)\sigma_--(\bar{v}-\bar{u})\sigma_+-i\Omega \sigma_3\right).
\end{equation}
Note that when $\alpha$ tends to zero the discontinuity disappears. Hence, when both parameters $\alpha$ and $\beta$ tend to zero the matrix ${\cal K}_D$ tends to the identity. Also note that ${\cal K}_D$ now depends on the fields $u$ and $v$. This is a common feature of the ${\cal K}_D$ matrix associated with this kind of defect.
In \cite{cz06NLS}, in order to find this matrix a modified Lax pair argument was used. Briefly, according to the idea of modified Lax pairs - first introduced in \cite{bcdr95} (see also \cite{bczlandau}) - two points $a<0$ and $b>0$ are introduced. They are the endpoints of two regions overlapping the defect, one on the left $R^-$, $-\infty<x<b$ and one on the right $R^+$, $a<x<\infty$. Relying on the explicit knowledge of the defect conditions, suitable Lax pairs $(U^-,V^-)$ and $(U^+,V^+)$ can be defined on each region. They provide both the equations of motions for the field $u$ and $v$ and the boundary conditions as a consequence of the zero curvature \eqref{zerocurvaturecondition}. In the overlapping interval $a<x<b$ the matrices $V^-$, $V^+$ must be related by the following gauge transformation (see \eqref{gaugetranformations})
\begin{equation}\label{constrainonK}
\partial_t {\cal K}_D=V^+(b,t){\cal K}_D-{\cal K}_DV^-(a,t).
\end{equation}
It is this expression that allows to find the matrix ${\cal K}_D$ \eqref{Kmatrix}. In this context, the matrix ${\cal K}_D$ is also called the B\"acklund matrix \cite{HabKun07}, which relates the eigenfunctions $\phi^-(v,\lambda)$ of the linear system
\begin{eqnarray*}
\partial_x \phi^-(x,t,\lambda)&=&U^-\phi^-(x,t,\lambda),\nonumber\\
\partial_t \phi^-(x,t,\lambda)&=&V^-\phi^-(x,t,\lambda),
\end{eqnarray*}
to the eigenfunctions $\phi^+(u,\lambda)$ of a similar linear system, as follows
$$\phi^+(u,\lambda)={\cal K}_D(u,v,\lambda)\phi^-(v,\lambda).$$

\section{The nonlinear Schr\"odinger model with an integrable boundary}
\label{thenewboundary}

\p In this section Sklyanin's formalism will be sketched and a new integrable boundary will be described. Consider the NLS field $u$ restricted to the positive $x$-axis\footnote{Note the difference with respect to the usual convention to confine the model to the negative $x$-axis}. This means that at $x=0$ a boundary term ${\bf B}$ is added and the full Lagrangian is
\begin{equation}\label{boundarylagrangian}
L=\int^{\infty}_0 dx\,{\cal
L}_u+{\bf B}.
\end{equation}
In \cite{Sklyanin88}, Sklyanin provided the first non trivial example of an integrable boundary for the NLS model. According to \cite{Sklyanin88} the term ${\bf B}$ is:
\begin{equation}\label{SklyaninBT}
{\bf B}=\gamma |u|^2,
\end{equation}
where $\gamma$ is a constant. Notice that the case $\gamma=0$ corresponds to the Neumann boundary condition and $\gamma=\infty$ corresponds to the Dirichlet boundary condition.
Using the boundary term \eqref{SklyaninBT}, the boundary condition at $x=0$ is:
\begin{equation}\label{bcSklyanin}
u_x=-\gamma u.
\end{equation}
In order to construct a generating functional for the conserved charges, a suitable monodromy matrix is needed. Sklyanin provides a generalisation of the monodromy matrix to the boundary case which is:
\begin{equation}\label{monodromymatrixSklyanin}
{\cal T}(\lambda)=T(\infty,0,t,\lambda)\,{\cal K}(\lambda)\,T^{-1}(\infty,0,t,-\lambda).
\end{equation}
The crucial step in proving that the conserved charges are in involution, relies on establishing the following relation
\begin{eqnarray}\label{rTboundaryrelation}
\{{\cal T}_1(\lambda), {\cal T}_2(\mu)\}&=&[r_{12}(\lambda-\mu),{\cal T}_1(\lambda)
{\cal T}_2(\mu)]\nonumber\\
&&-{\cal T}_2(\mu))\,r_{12}(\lambda+\mu)\,{\cal T}_1(\lambda)
+{\cal T}_1(\lambda)\,r_{12}(\lambda+\mu)\,{\cal T}_2(\mu).
\end{eqnarray}
The ${\cal K}$ matrix associated with the boundary \eqref{SklyaninBT} is:
\begin{equation}\label{boundaryKmatrix}
{\cal K}=\sigma_3-\frac{i\gamma}{\lambda}I,
\end{equation}
and ${\cal K}=I$ corresponds to the Dirichlet boundary condition.

\p As suggested in \cite{bu2008} (see also \cite{bajnok07}, \cite{cz2012}) it is possible to combine a defect with a boundary to create a new boundary. Imagine to put a defect in front of a boundary located at $x=0$. On the negative $x$-axis the field is $v$ and on the positive $x$-axis the field is $u$. Then assume that the field $v$ at the boundary satisfies the Dirichlet boundary condition, that is $v=0$ and $v_t=0$. Consider the defect conditions \eqref{defectconditions} with $v=v_t=0$ and take the sum and the difference of the two equations. It can be noticed that one of these equations provides an expression for the field $v_x$. Such an expression can be inserted into the other equation to provide a new boundary condition for the NLS field $u$ restricted to the positive half $x$-axis, which is
\begin{equation}\label{newBoundaryConditions}
u_x=\frac{iu_t}{2\Omega}-\frac{u\Omega}{2}+\frac{u|u|^2}{2\Omega}-\frac{u\beta^2}{2\Omega},\quad \Omega^2=(\alpha^2-|u|^2).
\end{equation}
The corresponding boundary term in the Lagrangian \eqref{boundarylagrangian} is:
\begin{equation}\label{newBoundaryterm}
{\bf B}=\frac{i\Omega}{2}\left(\frac{u_t}{u}-\frac{\bar{u}_t}{\bar{u}}\right)+\frac{\Omega^3}{3}+\Omega |u|^2-\Omega^3\beta^2.
\end{equation}
Notice the appearance of terms containing time derivatives in both the boundary term and the boundary condition. Also notice that by applying the same trick to the sine-Gordon model and the corresponding defect (type I \cite{bczlandau}), no new boundaries are obtained. The time derivative terms cancel out and in order to find new boundary it is necessary to use the type II defect \cite{cz2009} instead, which contains additional degrees of freedom, as shown in \cite{cz2012}.

\p In order to collect evidence to support the claim of integrability of the new boundary and eventually prove it, the corresponding ${\cal K}$ matrix is needed and the modified Lax pair technique will be used to find it. As mentioned in the previous section two regions $R^-$, $-\infty<x<b$ and $R^+$, $a<x<\infty$ overlap in a small interval $a<x<b$ around the boundary location $x=0$. Making use of the boundary condition \eqref{newBoundaryConditions} the Lax pair can be defined in each region, they are:
\begin{eqnarray*}
V^+(u)&=&V(u)+\theta(b-x)\left[\left(u_x-\frac{iu_t}{2\Omega}+\frac{u\Omega}{2}
-\frac{u|u|^2}{2\Omega}+\frac{u\beta^2}{2\Omega}\right)\sigma_-\right.\\
&&\left.-\left(\bar{u}_x+\frac{i\bar{u}_t}{2\Omega}+\frac{\bar{u}\Omega}{2}
-\frac{\bar{u}|u|^2}{2\Omega}+\frac{\bar{u}\beta^2}{2\Omega}\right)\sigma_+\right],\qquad
U^+(u)=U(u)\,\theta(x-b),
\end{eqnarray*}
and
\begin{eqnarray}\label{newboundaryLaxpair}
V^-(u)&=&V(u)+\theta(x-a)\left[\left(u_x+\frac{iu_t}{2\Omega}-\frac{u\Omega}{2}
+\frac{u|u|^2}{2\Omega}-\frac{u\beta^2}{2\Omega}\right)\sigma_-\right.\\\nonumber
&&\left.-
\left(\bar{u}_x-\frac{i\bar{u}_t}{2\Omega}-\frac{\bar{u}\Omega}{2}
+\frac{\bar{u}|u|^2}{2\Omega}-\frac{\bar{u}\beta^2}{2\Omega}\right)\sigma_+\right],
\qquad U^-(u)=U(u)\,\theta(a-x).
\end{eqnarray}
A few key observations are in order. The region $R^-$ will be regarded as a reflection of the region $R^+$ according to the following reflection principle
\begin{equation}\label{refprinciple}
u(x)=-u(a+b-x),
\end{equation}
where $x\in (-\infty,b),$ that is the region $R^-$. Notice that this reflection principle is different to the one used in \cite{bcdr95}. At first, it could seem a little bit weird, however it is perfectly justified. Consider the monodromy matrix that is required, it is \cite{bcdr95}:
\begin{equation}\label{monodromymatrixNBBCDR}
{\cal T}(\lambda)=T(\infty,b,t;\lambda)\,{\cal K}(\lambda)\,T(a, -\infty,t;\lambda).
\end{equation}
The idea is to use a suitable reflection principle in order to rewrite the element $T(a, -\infty,t;\lambda)$ in \eqref{monodromymatrixNBBCDR} defined over the region $R^-$ as an element over the region $R^+$ in such a way that \eqref{monodromymatrixNBBCDR} becomes \cite{Sklyanin88}
 \begin{equation}\label{monodromymatrixNBS}
{\cal T}(\lambda)=T(\infty,b,t;\lambda)\,{\cal K}(\lambda)\,T^{-1}(\infty,b,t;-\lambda).
\end{equation}
It is exactly the reflection principle \eqref{refprinciple} that allows to identity the two expressions.
In fact, notice that
$$
T^{-1}(\infty,b,t;-\lambda)=T(b,\infty, t;-\lambda)={\cal P}\exp\left(\int_\infty^b dz\, U(u(z),-\lambda)\right).
$$
Perform the change of variable $z= a+b-x$, then
$$
{\cal P}\exp\left(\int_\infty^b dz\, U(u(z),-\lambda)\right)={\cal P}\exp\left(\int_{-\infty}^a (-dx)\, U(u(a+b-x),-\lambda)\right).
$$
Finally, using the reflection principle \eqref{refprinciple}
$$U(u(a+b-x),-\lambda)=-U(u(x),\lambda),$$
which implies
$$
T^{-1}(\infty,b,t,-\lambda)=T(a,-\infty,t,\lambda).
$$
It is now possible to use the equation \eqref{constrainonK}, which also holds in the case of boundaries, in order to find an expression for the ${\cal K}$ matrix. The Lax pairs \eqref{newboundaryLaxpair} together with \eqref{refprinciple} are used. The reader should bear in mind that the final region of interest is the positive half $x$-axis. Notice that it is useful to use the following ansatz for the matrix ${\cal K}$
$${\cal K}=A+ B\sigma_3+ C\sigma_-+ D\sigma_+,$$
where $A$, $B$, $C$ and $D$ are coefficient that depend on the spectral parameter $\lambda$ and the field $u$. Also notice that the matrix ${\cal K}$ is expected to depend on the field $u$ as was the case for the defect situation. The expression \eqref{constrainonK}
provides the following constraints on these coefficients
\begin{eqnarray*}
A_t&=&-\lambda i(\bar{u}C +u D),\\
B_t&=&\left(\frac{iu_t}{2\Omega}-\frac{u\Omega}{2}
+\frac{u|u|^2}{2\Omega}-\frac{u\beta^2}{2\Omega}\right)D+\left(-\frac{i\bar{u}_t}{2\Omega}-\frac{\bar{u}\Omega}{2}
+\frac{\bar{u}|u|^2}{2\Omega}-\frac{\bar{u}\beta^2}{2\Omega}\right),\\
C_t&=&\left(-\frac{iu_t}{\Omega}+u\Omega
-\frac{u|u|^2}{\Omega}+\frac{u\beta^2}{\Omega}\right)B-2i\lambda u|+2i|u|^2C-\lambda^2iC,\\
D_t&=&\left(\frac{i\bar{u}_t}{\Omega}+\bar{u}\Omega
-\frac{\bar{u}|u|^2}{\Omega}+\frac{\bar{u}\beta^2}{\Omega}\right)B-2i\lambda \bar{u}|-2i|u|^2D+\lambda^2iD.
\end{eqnarray*}
Without loss of generality it is assumed that the coefficient $A$ is independent of the field $u$. A solution to the constraints is:
\begin{equation}\label{newboundaryKmatrix}
{\cal K}=\left(1-\frac{(\alpha^2+\beta^2)}{\lambda^2}\right)I+\frac{2}{\lambda}
\left(\bar{u}\sigma_+-u\sigma_--i \Omega\sigma_3\right).
\end{equation}
Notice that a similar calculation can be performed by using the boundary condition \eqref{bcSklyanin}. In that case the matrix \eqref{boundaryKmatrix} is recovered.

\subsection{On the integrability of the new boundary}
\label{sectionIntegrability}

\p The last step in the attempt to prove the integrability of the new boundary is to show that the boundary monodromy matrix \eqref{monodromymatrixNBS}, where $b$ is sent to zero - the defect location - satisfies expression \eqref{rTboundaryrelation} with the ${\cal K}$ matrix given by \eqref{newboundaryKmatrix}. Consider the left hand side of \eqref{rTboundaryrelation}. A little bit of algebra allows to rewrite it as follows
\begin{eqnarray}\label{PBFirstPart}
\{T_1,T_2\}({\cal K}T^{-1})_1({\cal K}T^{-1})_2+(T{\cal K})_1(T{\cal K})_2\{T^{-1}_1,T^{-1}_2\}
+({\cal K}T)_1\{T^{-1}_1, T_2\}({\cal K}T^{-1})_2\nonumber\\
+(T{\cal K})_2\{T_1, T^{-1}_2\}({\cal K}T^{-1})_1+T_1T_2\{{\cal K}_1, {\cal K}_2\}T^{-1}_1 T^{-1}_2,\phantom{********}
\end{eqnarray}
where a concise, nevertheless understandable, notation has been used.
The value of the Poisson brackets $\{T_1, T_2\}$ is given in \eqref{PBforT} and the Poisson bracket involving $T$ matrices can also be calculated (see \cite{FaddeevTakh86} where these kinds of computations are nicely explained). They are:
\begin{eqnarray*}
\{T^{-1}_1(x,y,-\lambda), T_2^{-1}(x,y,-\mu)\}&=&[r_{12}(\lambda-\mu),T^{-1}_1(x,y,-\lambda) T^{-1}_2(x,y,-\mu)],\\
\{T_1(x,y,\lambda),T^{-1}_2(x,y,-\mu)\}&=&T_1(x,y,\lambda)\,r_{12}\,(\lambda+\mu)T_2^{-1}(x,y,-\mu)\\
&&-T_2^{-1}(x,y,-\mu)\,r_{12}(\lambda+\mu)\,T_1(x,y,\lambda),\\
\{T_1^{-1}(x,y,-\lambda),T_2(x,y,\mu)\}&=&T_1^{-1}(x,y,-\lambda)\,r_{12}\,(\lambda+\mu)T_2(x,y,\mu)\\
&&-T_2(x,y,\mu)\,r_{12}(\lambda+\mu)\,T_1^{-1}(x,y,-\lambda),
\end{eqnarray*}
Then \eqref{PBFirstPart} becomes
\begin{eqnarray}\label{PBFirstPartBis}
&&[r_{12}(\lambda-\mu),{\cal T}_1(\lambda){\cal T}_2(\mu)]-{\cal T}_2(\mu)r_{12}(\lambda+\mu){\cal T}_1(\lambda)
+{\cal T}_1(\lambda)r_{12}(\lambda+\mu){\cal T}_2(\mu)\nonumber\\
&&+T_1(\lambda)T_2(\mu)\left[\,\,\{{\cal K}_1,{\cal K}_2\}-r_{12}(\lambda-\mu){\cal K}_1(\lambda){\cal K}_2(\mu)+{\cal K}_1(\lambda){\cal K}_2(\mu)r_{12}(\lambda-\mu)\right.\nonumber\\
&&\left.-{\cal K}_1(\lambda)r_{12}(\lambda+\mu) {\cal K}_2(\mu)
+{\cal K}_2(\mu)r_{12}(\lambda+\mu){\cal K}_1(\lambda)\,\,\right]T_1^{-1}(-\lambda)T_2^{-1}(-\mu).
\end{eqnarray}
The first line of \eqref{PBFirstPartBis} provides the terms on the right hand side of \eqref{rTboundaryrelation}. Hence, it must be
\begin{eqnarray}\label{PBInsidePart}
\{{\cal K}_1,{\cal K}_2\}&=&r_{12}(\lambda-\mu){\cal K}_1(\lambda){\cal K}_2(\mu)-{\cal K}_1(\lambda){\cal K}_2(\mu)r_{12}(\lambda-\mu)\nonumber\\
&&+{\cal K}_1(\lambda)r_{12}(\lambda+\mu) {\cal K}_2(\mu)
-{\cal K}_2(\mu)r_{12}(\lambda+\mu){\cal K}_1(\lambda).
\end{eqnarray}
Note that the ${\cal K}$ matrix does depend on the field $u$, hence its Poisson brackets are not automatically zero.

\p First consider the right hand side of \eqref{PBInsidePart}. Using the explicit expressions \eqref{rmatrix} for the $r$ matrix and \eqref{newboundaryKmatrix} for the ${\cal K}$ matrix, it is possible to compute
\begin{eqnarray}\label{intermidiatesteprhs}
r_{12}(\lambda-\mu){\cal K}_1(\lambda){\cal K}_2(\mu)-{\cal K}_1(\lambda){\cal K}_2(\mu)r_{12}(\lambda-\mu)\qquad\qquad\qquad\qquad\nonumber\\
=\frac{2}{\lambda\mu}\left(1+\frac{\alpha^2+\beta^2}{\lambda\mu}\right)\left[P,I\otimes(\bar{u}\sigma_+-u\sigma_--i\Omega\sigma_3)\right]
\nonumber\\
\nonumber\\
{\cal K}_1(\lambda)r_{12}(\lambda+\mu) {\cal K}_2(\mu)
-{\cal K}_2(\mu)r_{12}(\lambda+\mu){\cal K}_1(\lambda)\qquad\qquad\qquad\qquad\nonumber\\
=\frac{2}{\lambda\mu}\left(1-\frac{\alpha^2+\beta^2}{\lambda\mu}\right)\left[P,I\otimes(\bar{u}\sigma_+-u\sigma_--i\Omega\sigma_3)\right],
\end{eqnarray}
where property \eqref{permutationoperator} has been used.
Noticing that
$$\frac{P}{\lambda\mu}=\frac{r}{\mu}-\frac{r}{\lambda},$$
expressions \eqref{intermidiatesteprhs}, combine together, provide the right hand side of \eqref{PBInsidePart}, which becomes
\begin{equation}
\frac{4}{\lambda\mu}\left[P,I\otimes(\bar{u}\sigma_+-u\sigma_--i\Omega\sigma_3)\right]
=2\left[r_{12},{\cal K}_1(\lambda)+{\cal K}_2(\mu)\right].
\end{equation}

\p In order to calculate the left hand side of \eqref{PBInsidePart}, the standard definition for the Poisson brackets \eqref{PoissonBracketsbulk} needs to be slightly modified. In fact, consider the Lagrangian density \eqref{boundarylagrangian} with boundary term \eqref{newBoundaryterm}. Because of the presence of time derivatives, the canonical momenta conjugate to the fields $u$ and $\bar{u}$ are different with respect to the conjugate canonical momenta for the system without the new boundary. In fact, by definition, they are:
\begin{equation}\label{conjugatemomenta}
\frac{\delta{\cal L}}{\delta u_t}=i\frac{\bar{u}}{2}\theta(x)+i\frac{\Omega}{2u}\delta(x)=\pi\theta(x)+\rho\delta(x),\quad
\frac{\delta{\cal L}}{\delta \bar{u}_t}=-i\frac{u}{2}\theta(x)-i\frac{\Omega}{2\bar{u}}\delta(x)=\bar{\pi}\theta(x)+\bar{\rho}\delta(x).
\end{equation}
This strongly suggests the need to modify the Poisson brackets in order to take into account relations \eqref{conjugatemomenta} at the boundary location. Note how in \cite{cz2009} a similar observation concerning the conjugate canonical momenta was used in order to discuss second class constraints in the context of defects.
Then, by definition, the canonical Poisson brackets are:
\begin{eqnarray*}
\{A, B\}&=&\int_{-\infty}^{\infty}dz \left\{ \left(\frac{\delta A}{\delta \pi(z)}\frac{\delta B}{\delta u(z)}-\frac{\delta A}{\delta u(z)}\frac{\delta B}{\delta \pi(z)}+\frac{\delta A}{\delta \bar{\pi}(z)}\frac{\delta B}{\delta \bar{u}(z)}-\frac{\delta A}{\delta \bar{u}(z)} \frac{\delta B}{\delta \bar{\pi}(z)}\right)\theta(z)\right.\nonumber\\
&&\left.+\left(\frac{\delta A}{\delta \rho(z)}\frac{\delta B}{\delta u(z)}-\frac{\delta A}{\delta u(z)} \frac{\delta B}{\delta \rho(z)}+\frac{\delta A}{\delta \bar{\rho}(z)}\frac{\delta B}{\delta \bar{u}(z)}-\frac{\delta A}{\delta \bar{u}(z)} \frac{\delta B}{\delta \bar{\rho}(z)}\right)\delta(z)\right\}.
\end{eqnarray*}
After a little bit of algebra they can be expressed in terms of the fields $u$ and $\bar{u}$, since  $\pi=i\bar{u}/2$, $\bar{\pi}=-i u/2$, $\rho=i\Omega/2u$ and $\bar{\rho}=-i\Omega/2\bar{u}$. Hence, the Poisson brackets become
\begin{equation}\label{newPoissonBrackets}
\{A, B\}=i\int_{0}^{\infty}dz \left(\frac{\delta A}{\delta u(z)}\frac{\delta B}{\delta \bar{u}(z)}-\frac{\delta A}{\delta \bar{u}(z)}\frac{\delta B}{\delta u(z)}\right)
-2i\Omega\left(\frac{\partial A}{\partial u}\frac{\partial B}{\partial \bar{u}}-\frac{\partial A}{\partial \bar{u} } \frac{\partial B}{\partial u}\right),
\end{equation}
where an overall factor of four has been removed. Notice the difference with respect to the Poisson brackets \eqref{PoissonBracketsbulk}. Clearly, the computations performed previously involving the Poisson brackets for the $T$ matrices are still valid. In addition, the Poisson brackets for the ${\cal K}$ matrices can now be calculated. Since the ${\cal K}$ matrix holds at the boundary location, only the second term in the Poisson brackets \eqref{newPoissonBrackets} is used. Hence
$$
\{{\cal K}(\lambda), {\cal K}(\mu)\}=-\frac{4i\Omega}{\lambda\mu}\left(2(\sigma_+\otimes\sigma_--\sigma_-\otimes\sigma_+)
+\frac{i\sigma_3}{\Omega}\otimes(u\sigma_-+\bar{u}\sigma_+)-(u\sigma_-+\bar{u}\sigma_+)\otimes\frac{i\sigma_3}{\Omega}\right).
$$
Using the facts that
$$2(\sigma_+\otimes\sigma_--\sigma_-\otimes\sigma_+)=[P,I\otimes \sigma_3]$$
and
$$(u\sigma_-+\bar{u}\sigma_+)\otimes \sigma_3-\sigma_3\otimes(u\sigma_-+\bar{u}\sigma_+)=[I\otimes (\bar{u}\sigma_+-u\sigma_-),P],$$
the Poisson brackets of the boundary ${\cal K}$ matrix become
\begin{equation}
\{{\cal K}(\lambda), {\cal K}(\mu)\}=2\left[r_{12},{\cal K}_1(\lambda)+{\cal K}_2(\mu)\right],
\end{equation}
as expected. As a consequence, expression \eqref{PBInsidePart} becomes an identity and
integrability is proved.

\p As a final note, consider the quantity $\Omega$ defined in \eqref{newBoundaryConditions}. When $\Omega$ is equal to zero the NLS fields at the boundary acquires a fixed value. In particular, when the parameter $\alpha$ goes to zero the NLS fields at the boundary reduce to zero and the new boundary turns into the Dirichlet boundary. This makes perfectly sense since, in this case, the defect used to create the new boundary disappears. Hence the Poisson brackets \eqref{newPoissonBrackets} reduce to the more familiar brackets used in the case of the NLS field restricted to a half line with, for instance, Dirichlet boundary conditions.

\subsection{Conserved charges}

\p In this section it will be shown how the conserved charges can be obtained by using a suitable generating functional. In order to do so a brief summary of the procedure used in the bulk will be also presented. Consider first the transition matrix $T(L, -L, \lambda)$ in the bulk. Note that this matrix is defined on an interval. Periodic boundary conditions are chosen on the fields with period $2L.$ Results for the NLS on the full $x-$axis are obtained by simply sending $L$ to $\infty$ in the final results. In that case the fields and their derivatives are assumed to satisfy Schwartz boundary conditions at infinity (see section \ref{section2}). It can be proved \cite{FaddeevTakh86} that the transition matrix has the following asymptotic expansion for large real $\lambda$
\begin{equation}\label{monodromymatrixExpansion}
T(L, -L, \lambda)=(I+W(L,\lambda))\exp Z(L, -L,\lambda)(I+W(-L,\lambda))^{-1},
\end{equation}
with
\begin{eqnarray*}
W(L,\lambda)&=&\sum_{n=1}^{\infty}\frac{W_n(L)}{\lambda^n}\,O(|\lambda|^{-\infty}),\\
Z(L,-L,\lambda)&=&-iL\lambda\sigma_3-i\sum_{n=1}^{\infty}\frac{Z_n(L,-L)}{\lambda^n}\,O(|\lambda|^{-\infty}).
\end{eqnarray*}
The series coefficients are completely determined. They are:
\begin{equation}\label{Zexpansion}
Z(L,-L)=\left(
    \begin{array}{cc}
      z_n(L,-L) & 0 \\
      0 & -\bar{z}_n(L,-L) \\
    \end{array}
  \right),\quad z_n=\int^L_{-L}\bar{u}(z)w_n(z)\,dz,
\end{equation}
and
\begin{equation}\label{Wexpansion}
W(L)=\left(
         \begin{array}{cc}
           0 & \bar{w}_n(L) \\
           -w_n(L) & 0 \\
         \end{array}
       \right),\quad w_1=u,\quad w_{n+1}=-i(w_n)_x-\bar{u}\sum_{k=1}^{n-1}\,w_k w_{n-k}.
\end{equation}
The generating functional $Q(\lambda)$, that is the trace of the monodromy matrix, reduces to the following expression
\begin{equation}
Q(\lambda)=\mbox{Tr} (\exp Z(L, -L,\lambda))=2\cos (iL\lambda +i z_n)+O(|\lambda|^{-\infty}),\quad \lambda \rightarrow \pm i\infty,
\end{equation}
where the periodic boundary conditions have been used.
Without loss of generality consider the terms for which $\lambda \rightarrow i\infty$, then
\begin{equation}
\ln(\mbox{Tr} \,T(L, -L,\lambda))=-i \left(\frac{z_1}{\lambda}+\frac{z_2}{\lambda^2}+\frac{z_3}{\lambda^3}\right)+O(|\lambda|^{-4}),\quad \lambda \rightarrow i\infty,
\end{equation}
where the first three orders in $1/\lambda$ have only been considered. This expansion provides the conserved charges in the bulk. Explicitly, the first three conserved charges are:
$$I_1=-i\int^L_{-L}|u|^2 \,dz,\quad I_2=-\int^L_{-L}\bar{u}u_x \,dz,\quad I_3=i\int^L_{-L}(\bar{u}u_{xx}+|u|^4)\, dz.$$
They correspond - up to an overall factor of $-i$ - to the `probability' or `number', momentum and energy, respectively.

\p In the presence of a boundary the situation is slightly different. Consider the NLS model on the interval $0\leq x \leq L$. According to \cite{Sklyanin88} the generating functional is defined as follows
\begin{equation}
{\cal Q}(\lambda)=\mbox{Tr}\left({\cal \hat{K}}(L;\lambda){\cal T}(L,0;\lambda)\right)=
\mbox{Tr}\left({\cal \hat{K}}(L;\lambda)T(L,0;\lambda)\,{\cal K}(0;\lambda)\,T^{-1}(L,0;-\lambda)\right).
\end{equation}
The matrices ${\cal \hat{K}}$ and ${\cal K}$ take into account the specific boundary at $L$ and $0$, respectively. In the present case the matrix
${\cal \hat{K}}$ is set equal to the identity - Dirichlet boundary conditions\footnote{Note that this choice for the boundary condition at $L$ does not interfere with the final choice of Schwartz boundary conditions when $L$ is sent to $\infty$. } - while the matrix ${\cal K}$ is given by \eqref{newboundaryKmatrix}. In the end, the generating functional is provided by the trace of the monodromy matrix \eqref{monodromymatrixNBS} restricted to the interval $0\leq x \leq L$. Such a monodromy matrix can be written as follows
\begin{eqnarray*}
{\cal T}(L, 0; \lambda)&=&(I+W(L,\lambda))\exp Z(L, 0,\lambda)(I+W(0,\lambda))^{-1}{\cal K}(0,\lambda)\\
&&\cdot(I+W(0,-\lambda))\exp (-Z(L, 0,-\lambda))(I+W(L,-\lambda))^{-1},
\end{eqnarray*}
where \eqref{monodromymatrixExpansion} has been used. Hence
\begin{equation*}
{\cal Q}(\lambda)=\mbox{Tr}\left(\exp (Z(L, 0,\lambda)-Z(L, 0,-\lambda))(I+W(0,\lambda))^{-1}{\cal K}(0,\lambda)(I+W(0,-\lambda))\right).
\end{equation*}
The last group of terms can be rewritten as an asymptotic expansion for large values of $\lambda$ as follows
\begin{equation}\label{Hexpansion}
(I+W(0,\lambda))^{-1}{\cal K}(0,\lambda)(I+W(0,-\lambda))=\sum_{n=0}^{\infty}\frac{H_n(0)}{\lambda^n}+O(|\lambda|^{-\infty})
\end{equation}
with
\begin{equation}\label{hcoefficients}
H_n(0)=\left(
  \begin{array}{cc}
    h_n & \hat{h}_n \\
    \tilde{h}_n & \bar{h}_n \\
  \end{array}
\right),\quad h_0=1,\quad h_1=-2i\Omega,\quad h_2=-(\alpha^2+\beta^2)+2|u|^2,\quad h_3=2iu_x\bar{u},\quad \dots,
\end{equation}
where the matrix \eqref{newboundaryKmatrix} has been used. It should bear in mind that only the diagonal terms of the matrix $H_n$ are relevant. Finally
\begin{equation}\label{boundarygeneratingfunctional}
\ln(\mbox{Tr} \,{\cal T}(L, 0, \lambda))=-i\sum_{n=1}^\infty \left(1-(-1)^n\right)\frac{z_n}{\lambda^n}+
\ln\left(\sum_{n=1}^\infty\frac{h_n}{\lambda^n}\right)+O(|\lambda|^{-\infty}),\quad \lambda \rightarrow i\infty.
\end{equation}
Note that the even conserved charges disappear. Up to an overall factor of $-2i$, the first two conserved charges to survive, that is the `number' and the energy, are modified as follows
\begin{equation}\label{modifiedconservedcharges13}
\left.\hat{I}_1=-2i\int^L_{0}|u|^2 \,dz-2i\Omega|_{x=0},\quad \hat{I}_3=2i\int^L_{0}(|u|^4-\bar{u}_xu_x)\, dz+\left(-2i\Omega\beta^2+2i\Omega|u|^2+\frac{2}{3}i\Omega^3\right)\right|_{x=0},
\end{equation}
where the logarithm expansion up to order $1/\lambda^3$ has been used.

\p It can be easily verified that $\hat{I}_1$ and $\hat{I}_3$ coincide with the conserved charges that can be obtained starting with the bulk charge density and using the boundary conditions \eqref{newBoundaryConditions}, as first shown in \cite{gz1994}. In fact, consider the charges $I_1$ on the interval $0\leq x\leq L$, that is
$$I_1=-2i\int_0^L |u|^2\,dz.$$
Its time derivative is:
$$\partial_t I_1=-2i\int_0^L (u_t\bar{u}+u\bar{u}_t)\,dz=-2(u_x\bar{u}-u\bar{u}_x)|_{x=0},$$
where the equation of motion \eqref{NLSequation} has been used. Then, using the boundary condition \eqref{newBoundaryConditions} this expression can be rewritten as follows
$$\partial_t I_1=2i\partial_t \Omega|_{x=0}.$$
Hence, the modified conserved charge $\hat{I}_1$ is:
$$\hat{I}_1=-2i\int^L_{0}|u|^2 \,dz-2i\Omega|_{x=0}.$$
Similarly
$$I_3=2i\int_0^L (|u|^4+u_{xx}\bar{u})\,dz=2i\int_0^L (|u|^4-u_{x}\bar{u}_x)\,dz-2iu_x\bar{u}|_{x=0},$$
and its time derivative is:
$$\partial_t I_3=2i(u_t\bar{u}_x+u_x\bar{u}_t)|_{x=0}-2i\partial_t(u_x\bar{u})|_{x=0}.$$
Using the boundary conditions
$$\left.\partial_t I_3=\partial_t\left(-2i\Omega|u|^2-i\frac{2}{3}\Omega^3+2i\beta^2\Omega-2iu_x\bar{u}\right)\right|_{x=0}$$
Hence the modified conserved charge $\hat{I}_3$ is:
$$\hat{I}_3=2i\int^L_{0}(|u|^4-\bar{u}_xu_x)\, dz+\left.\left(-2i\Omega\beta^2+2i\Omega|u|^2+\frac{2}{3}i\Omega^3\right)\right|_{x=0}.$$
The conserved charges \eqref{modifiedconservedcharges13} obtained by expanding the trace of the monodromy matrix have been recovered.
This procedure can be easily extended to the higher order conserved charges, though the calculations are much heavier. As an example, the result for the first non trivial charge, $\hat{I}_5$, is presented in appendix \ref{appendixA}.

\section{Conclusions}

\p In this article the possibility to combine a boundary with a defect in order to create a new integrable boundary has been explored, classically, in the context of the NLS model. Using a Lagrangian approach a new boundary potential has been obtained by combining a Dirichlet boundary with a type I defect \cite{cz06NLS}. Unlike previous examples, the new boundary does not have any additional degrees of freedom and the boundary conditions are characterized by the presence of time derivatives of the NLS fields. Clearly, these can be eliminated using the equation of motion. However, the price to pay for doing so is the introduction of higher order space derivatives. The modified Lax pair technique has been used to find the associated ${\cal K}$ matrix, which, as expected, does depend on the NLS fields. The present case provides the first example of a non-constant boundary ${\cal K}$ matrix without additional degrees of freedom in the context of the NLS model. Then, the integrability of the new boundary has been proved via the classical $r$ matrix. In order to do so, modified Poisson brackets has been constructed. These take into account the non trivial coupling of the NLS fields at the boundary. Finally, the generating functional has been used to derived the first three conserved charges and these results have been found to agree with the conserved charges obtained by using a Lagrangian approach, that is by using simply the boundary conditions instead of the boundary ${\cal K}$ matrix.

\p The present investigation also offers the opportunity to prove the integrability of the type I defect. In fact, in order to prove that
\begin{equation}\label{integrabilitycondition}
\{{\cal Q}_D(\lambda), {\cal Q}_D(\mu)\}=0,
\end{equation}
it is necessary to compute the Poisson brackets of the defect ${\cal K}_D$ matrices.
The procedure is very similar to the one adopted in this article for the boundary case. The key point is the modification of the Poisson brackets as explained in section \ref{sectionIntegrability}. In the case of the type I defect associated with the defect ${\cal K}_D$ matrix \eqref{Kmatrix}, the Poisson brackets are taken to be
\begin{eqnarray*}
\{A, B\}&=&i\int_{0}^{\infty}dz \left(\frac{\delta A}{\delta u(z)}\frac{\delta B}{\delta \bar{u}(z)}-\frac{\delta A}{\delta \bar{u}(z)}\frac{\delta B}{\delta u(z)}\right)
-2i\Omega\left(\frac{\partial A}{\partial u}\frac{\partial B}{\partial \bar{u}}-\frac{\partial A}{\partial \bar{u}} \frac{\partial B}{\partial u}\right)\\
&=&i\int_{-\infty}^{0}dz \left(\frac{\delta A}{\delta v(z)}\frac{\delta B}{\delta \bar{v}(z)}-\frac{\delta A}{\delta \bar{v}(z)}\frac{\delta B}{\delta v(z)}\right)
-2i\Omega\left(\frac{\partial A}{\partial v}\frac{\partial B}{\partial \bar{v}}-\frac{\partial A}{\partial \bar{v}} \frac{\partial B}{\partial v}\right).
\end{eqnarray*}
Using this definition for the Poisson brackets, it is found that
$$
\{{\cal K}_D(\lambda), {\cal K}_D(\mu)\}=2\left[r_{12},{\cal K}_D(\lambda)\otimes 1+1\otimes{\cal K}_D(\mu)\right]
$$
which suffices to prove \eqref{integrabilitycondition}.
Clearly, the procedure explained in the present article can be extended to other models in order to prove the integrability of the type I and type II defects.

\p In this article soliton preserving (SP) boundary conditions have been considered. It would be interesting to see whether the dressing operating here can be extended to create new, soliton non preserving (SNP) boundary conditions. In this case the generating functional is going to change and this will affect the conserved charges. However, first it should be clarify whether the reflection principle used in section \ref{thenewboundary} for finding the ${\cal K}$ matrix still holds or should be modified.

\p The NLS model can be generalised to multi component complex scalar fields, the so called vector NLS models \cite{FaddeevTakh86}. In recent year these models have been the subject of several investigations in the context of defects \cite{vectorNLSdefects} and boundaries \cite{vectorNLSboundary} (see also \cite{dfr2008} and references therein). However, to the author's knowledge, no type I-like defects are known. The author believe strongly that vector NLS models can support such defects. It will be interesting to find them and then investigate their dressing of known boundaries.

\vskip .5cm
\p {\large \bf Acknowledgements}\\

\p I would like to thanks Ed Corrigan for reading the manuscript and providing useful suggestions.

\appendix
\section{Appendix: $\hat{I}_5$ charge}
\label{appendixA}

\p In this appendix the modified conserved charge $\hat{I}_5$ is derived. Consider the charges $I_5$ on the interval $0\leq x\leq L$. Using formulae \eqref{Zexpansion}, \eqref{Wexpansion} it reads
\begin{eqnarray*}
I_5&=&-2i\int_0^L \left(\bar{u}u_{xxxx}+6u_{xx}\bar{u}|u|^2+\bar{u}_{xx}u|u|^2+5u_x^2\bar{u}^2+6u_x\bar{u}_x|x|^2+2|u|^6\right)dz\\
&=&-2i\int_0^L \left(\bar{u}_{xx}u_{xx}-8u_{x}\bar{u}_x|u|^2-\bar{u}_x^2u^2-u_x^2\bar{u}^2+2|u|^6\right)dz\\
&&+2i\left(\bar{u}u_{xxx}-\bar{u}_xu_{xx}+6u_x\bar{u}|u|^2+\bar{u}_xu|u|^2\right)
\end{eqnarray*}
Using the equation of motion \eqref{NLSequation} its time derivative can be written as follows:
\begin{eqnarray*}
\partial_t I_5&=&2i \left(\bar{u}_{xt}u_{xx}+\bar{u}_{xx}u_{xt}-4u_x\bar{u}_t|u|^2-4u_t\bar{u}_x|u|^2\right)|_{x=0}\\
&&-4\left(\bar{u}_xu|u|^4-u_x\bar{u}|u|^4+u_xu\bar{u}_x^2-\bar{u}_x\bar{u}u_x^2\right)|_{x=0}\\
&&+\partial_t\left(\bar{u}u_{xxx}-\bar{u}_xu_{xx}+6u_x\bar{u}|u|^2+\bar{u}_xu|u|^2\right)|_{x=0}.
\end{eqnarray*}
Then, using the boundary condition \eqref{newBoundaryConditions} and fiddling with the algebra, this expression can be rewritten as a total time derivative
\begin{eqnarray*}
\partial_t I_5&=&2i\,\partial_t\left(\bar{u}u_{xxx}-\bar{u}_xu_{xx}+5u_x\bar{u}|u|^2-2\bar{u}_xu_x\Omega-
2(\bar{u}_xu+u_x\bar{u})(\Omega^2+\beta^2)\right)|_{x=0}\\
&&2i\,\partial_t\left.\left(\Omega|u|^4-4\Omega|u|^2\beta^2+\Omega\beta^4-2\Omega^3\beta^2+\frac{\Omega^4}{5}\right)\right|_{x=0}.
\end{eqnarray*}
Hence, the modified conserved charge is:
\begin{eqnarray}\label{I5charge1}
\hat{I}_5&=&-2i\int_0^L\left(\bar{u}u_{xxxx}+6u_{xx}\bar{u}|u|^2+\bar{u}_{xx}u|u|^2+5u_x^2\bar{u}^2+6u_x\bar{u}_x|x|^2+2|u|^6\right)dz\nonumber\\
&&-2i\,\left(\bar{u}u_{xxx}-\bar{u}_xu_{xx}+5u_x\bar{u}|u|^2-2\bar{u}_xu_x\Omega-
2(\bar{u}_xu+u_x\bar{u})(\Omega^2+\beta^2)\right)|_{x=0}\nonumber\\
&&-2i\,\left.\left(\Omega|u|^4-4\Omega|u|^2\beta^2+\Omega\beta^4-2\Omega^3\beta^2+\frac{\Omega^4}{5}\right)\right|_{x=0}.
\end{eqnarray}
An identical result is obtained by using the generating functional \eqref{boundarygeneratingfunctional}. In this case a lengthy calculation leads to the following expression
\begin{eqnarray}\label{I5charge2}
\hat{I}_5&=&-2i\int_0^L\left(\bar{u}u_{xxxx}+6u_{xx}\bar{u}|u|^2+\bar{u}_{xx}u|u|^2+5u_x^2\bar{u}^2+6u_x\bar{u}_x|x|^2+2|u|^6\right)dz\nonumber\\
&&-2i\,\left(\bar{u}u_{xxx}-\bar{u}_xu_{xx}+6\bar{u}u_x|u|^2+\bar{u}_xu|u|^2-(\bar{u}_xu+\bar{u}u_x)
(\alpha^2+\beta^2)-2\Omega\bar{u}_xu_x\right)|_{x=0}\nonumber\\
&&-2i\,\left.\left(8\Omega^3|u|^2+6\Omega|u|^4-(4\Omega^3+6\Omega|u|^2)(\alpha^2+\beta^2)
+\Omega(\alpha^2+\beta^2)^2+\frac{16\,\Omega^5}{5}\right)\right|_{x=0}.\nonumber\\
\end{eqnarray}
where the coefficients $h_i$ \eqref{hcoefficients} of the \eqref{Hexpansion} expansion have been used, together with
\begin{eqnarray*}
h_4&=&-2|u|^4+2|u|^2(\alpha^2+\beta^2)+4\bar{u}u_x\Omega\\
h_5&=&2i\left(u_{xx}\bar{u}_x-u_{xxx}\bar{u}-4\bar{u}u_x|u|^2+\bar{u}_xu(\alpha^2+\beta^2)-\bar{u}_xu|u|^2+2\bar{u}_xu_{x}\Omega\right).
\end{eqnarray*}
It can be easily verified that expressions \eqref{I5charge1} and \eqref{I5charge2} are identical provided the relation $\alpha^2=\Omega^2+|u|^2$
is used.

\end{document}